\documentstyle[proceedings]{crckapb}
\input epsf
\begin{opening}
\title{Bulge $\delta$ Scuti Stars in the MACHO Database}
\end{opening}
\begin{document}
%
%
%
%
%

\newcounter{authornum}
\newcounter{affilnum}
\newsavebox{\authorbox}
\newsavebox{\authorsave}
\newsavebox{\affilbox}
\newsavebox{\affilsave}

%
%

\newcommand{\super}[1]{$\strut^{\rm #1}$}


\newenvironment{poster}[1]
{
\setcounter{authornum}{0}
\setcounter{affilnum}{0}
\begin{flushleft}
\noindent{\Large\textbf{#1}}
\end{flushleft}
}
{
\vfill
\newpage
}
\newenvironment{authors}
{
\sbox{\authorbox}{\strut}
\sbox{\authorsave}{\strut}
\begin{flushleft}
}
{
\usebox{\authorsave}
\end{flushleft}
}

\renewcommand{\author}[1]
{%
#1,
\stepcounter{authornum}}

\newcommand{\iauthor}[2]
{%
\author{#2\super{#1}}%
\stepcounter{authornum}}

\newenvironment{affiliations}
{
\sbox{\affilbox}{\strut} 
\small\it
\begin{flushleft}
}
{
\end{flushleft}
}

\newcommand{\affiliation}[1]
{%
#1;
\stepcounter{affilnum}}

\newcommand{\iaffiliation}[2]
{%
\affiliation{\super{#1}#2}%
\stepcounter{affilnum}}

\newenvironment{references}
{
\small
\begin{flushleft}
{\sc\lowercase{references}}\\
\vspace{1mm}
}
{
\end{flushleft}
}

\newcommand{\reference}[1]
{%
#1\\
}

\setlength{\textwidth}{125mm}
\setlength{\textheight}{190mm}
\setlength{\oddsidemargin}{.2in}
\setlength{\evensidemargin}{.2in}
\setlength{\marginparwidth}{1in}
\setlength{\marginparsep}{10pt}
\setlength{\topmargin}{-0.6mm}
\setlength{\headheight}{12pt}
\setlength{\headsep}{6mm}
\setlength{\footskip}{39pt}

\renewcommand{\chapter}{}


\newcommand{\ion}[2]{\hbox{#1\hspace{0.15em}{\small\rm{#2}}}}

%
%
\newcommand{\finish}{\end{document}}

 \begin{authors}
\iauthor{1}{D. MINNITI}
\iauthor{1}{C. ALCOCK}
\iauthor{1}{D.R. ALVES}
\iauthor{2}{T.S AXELROD}
\iauthor{3}{A.C BECKER}
\iauthor{4}{D.P. BENNETT}
\iauthor{1}{K.H. COOK}
\iauthor{2}{K.C. FREEMAN}
\iauthor{5}{K. GRIEST}
\iauthor{5}{M.J. LEHNER}
\iauthor{1}{S.L. MARSHALL}
\iauthor{2}{B.A. PETERSON}
\iauthor{6}{P.J. QUINN}
\iauthor{3}{M.R. PRATT}
\iauthor{2}{A.W. RODGERS}
\iauthor{3}{C.W. STUBBS}
\iauthor{7}{W. SUTHERLAND}
\iauthor{3}{A. TOMANEY}
\iauthor{5}{T. VANDEHEI}
\iauthor{8}{D. WELCH}
\end{authors}
\runningtitle{Bulge $\delta$ Scuti Stars}
\begin{affiliations}
\iaffiliation{1}{LLNL}
\iaffiliation{2}{MSSSO}
\iaffiliation{3}{U. Washington}
\iaffiliation{4}{Notre Dame}
\iaffiliation{5}{UC San Diego}
\iaffiliation{6}{ESO}
\iaffiliation{7}{Oxford U.}
\iaffiliation{8}{McMaster U.}
\end{affiliations}

\section{Introduction}

{\it We describe the search for $\delta$ Scuti stars in the MACHO database
of bulge fields. Concentrating on a sample of high amplitude $\delta$ Scutis,
we examine the light curves and pulsation modes. We also discuss their
spatial distribution and
evolutionary status using mean colors and absolute magnitudes.}\\

The 61" telescope at Mt. Stromlo Observatory observes once a night several
bulge or LMC fields in two bandpasses. While the MACHO system is optimized
for microlensing, it has also had great success detecting variable stars
(Cook et al. 1995).  We have searched for $\delta$ Scuti stars in the bulge
fields (these variable stars are beyond the detection limit in the LMC fields).
The $\delta$ Scutis have very short periods, typically between 1 hr and 6 hr
(e.g. Breger 1995, Rodriguez et al. 1994), 
which may be challenging using a once-a-day observing routine. Amazingly enough,
the light curves of $\delta$ Scuti stars phase very well for periods as short as
$0.08$ days. Because subsequent observations are separated by 5-15 pulsation
periods, the phases are randomized, giving a rather uniform light curve coverage.
Also, because we have so many datapoints per star ($\sim 500$ over four seasons),
aliasing is not a significant problem. The quality of the blue and red instrumental
light curves (Figure 1) say more than a thousand words...


The $\delta$~Scuti stars are selected using the
period-amplitude diagram for all stars with $P < 0.20^d$,
and amplitudes $A_V > 0.1$ mag in the MACHO blue band,
taking for comparison the catalogue of Rodriguez et al. (1994).
We demand that $P_V = P_R$ to within 2\%.
The $\delta P$ criterion implies that we keep mainly the variables that have
good quality light curves, and that are not multi-mode pulsators.
The lower cut in the amplitudes is imposed to secure good data, even though
small amplitude $\delta$~Scuti stars will also be neglected.
We also demand that $A_R < 0.8 A_V$, in order to 
eliminate eclipsing binaries which have $A_V \approx A_R$.
Finally, the identification
of these large amplitude $\delta$~Scuti candidates is confirmed by
visual inspection of the light curves.

\vskip 9.08truecm
\begin{minipage}{10cm}
\includegraphics{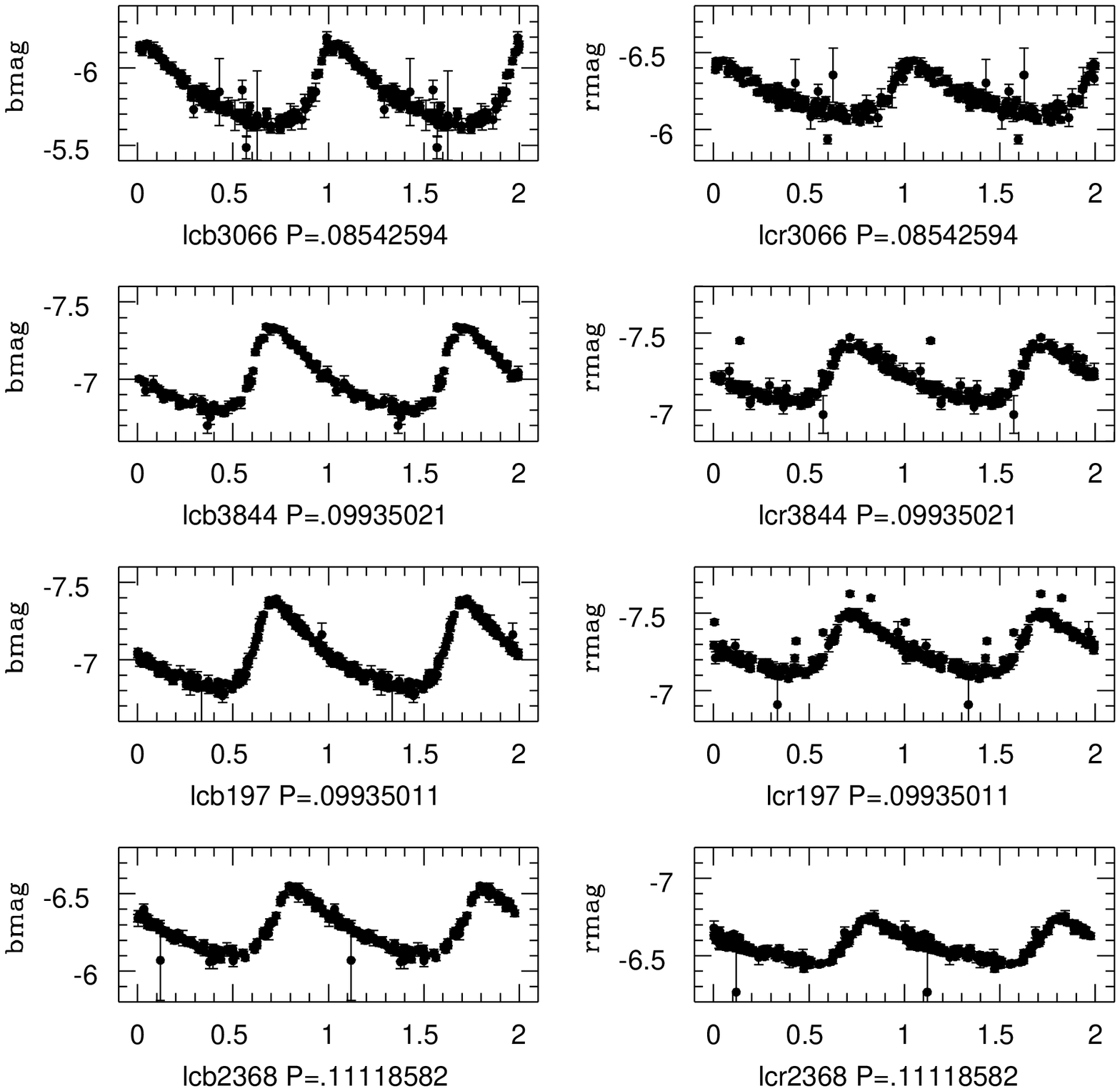}
\end{minipage}
\smallskip{}
\begin{center}
{\small Fig. 1 - Typical blue (left panels) and red (right panels) $\delta$~Scuti light curves.}
\end{center}


Thus, we do not consider the low amplitude,
the short period, and most double or multi-mode $\delta$~Scuti stars.
Even though the sample is incomplete, it is
adequate to examine the extension of the existing
P-L-Z relations into the metal-rich
domain, and to determine the evolutionary status of these stars in the bulge.

\section{Light Curves: Fundamental, 1st Overtone, or 2nd Overtone?}

The light curve shapes of the $\delta$ Scuti stars in our sample
are similar to short-period versions of RR Lyrae type ab and type c variables.
Theoretical light curves for $\delta$~Scuti stars pulsating in the fundamental,
first overtone, and second overtone modes have been recently computed by
Bono et al. (1997). In particular, they predict for the first time the
occurrence of stable second overtone pulsators.
While a more detailed exploration of the full parameter space
is needed, these models (Bono et al. 1997, their Figures 3--7) 
reproduce the observed light curves of our
sample very well.

\vskip 3.08truecm
\begin{minipage}{3cm}
\includegraphics{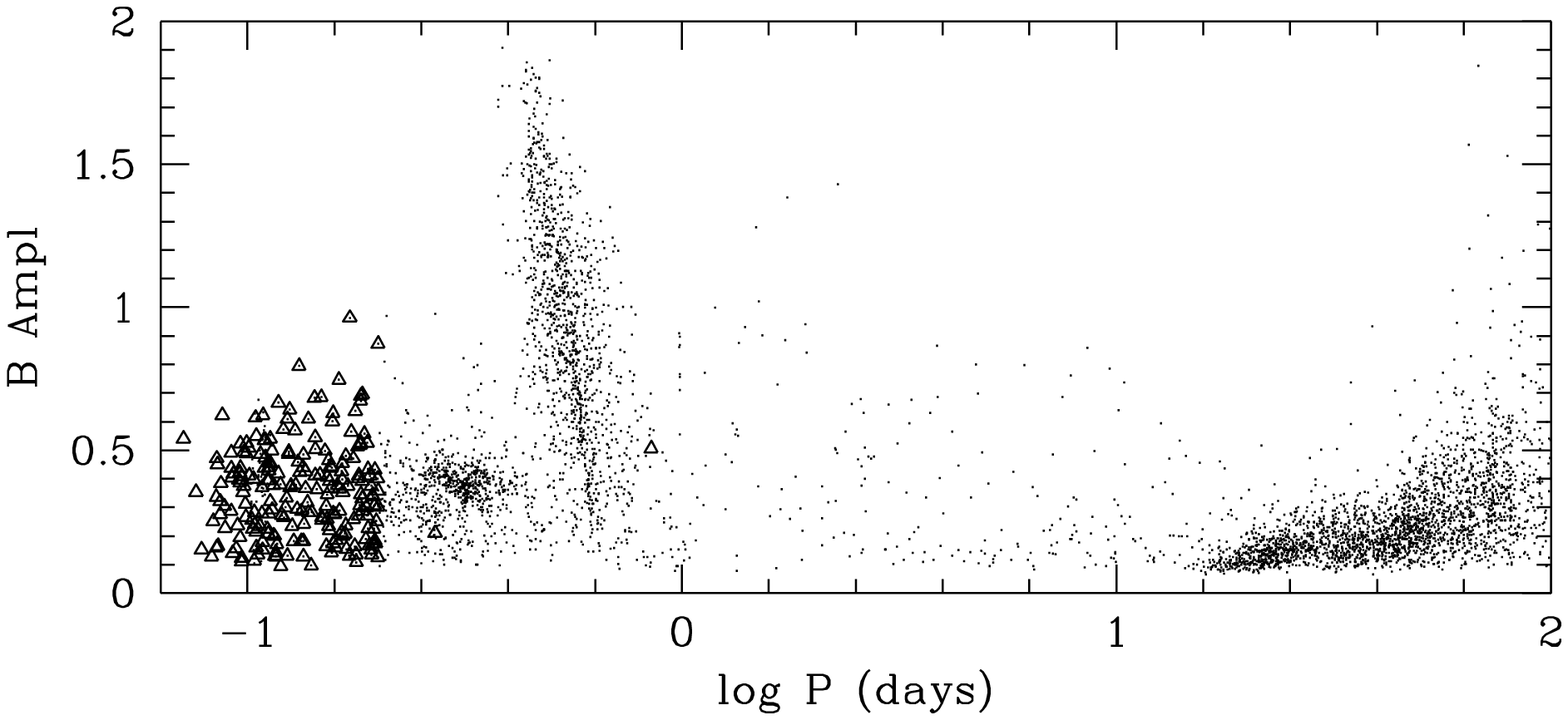}
\end{minipage}
\smallskip{}
\begin{center}
{\small Fig. 2 - Period-amplitude diagram for pulsating bulge
variables with P$<$ 100$^d$.
The $\delta$ Scuti are plotted with triangles. RR Lyrae stars type ab, c, and 
e are also seen at $log~P \approx -0.25$, $-0.5$, and $-0.6$ days, respectively.
}
\end{center}


Few fundamental $\delta$~Scuti
pulsators would be included in the sample, since
these would have $P > 0.12^d$ and $A \leq 0.1$ mag.
Based on these periods and amplitudes, it appears that the
majority of the sample stars would be second overtone pulsators,
with a few first overtone pulsators.

Bono et al. (1997) find that, contrary to different RR Lyrae pulsators,
the second overtone $\delta$~Scuti pulsators show larger amplitudes, in spite of
having smaller periods than the first overtone or the fundamental
pulsators. This is in direct contradiction with the assumption
that, like RR Lyrae and Cepheid stars,
the higher modes have lower amplitudes.
Furthermore, the theoretical light curves of Bono et al. (1997) show that
fundamental $\delta$~Scuti pulsators should have sinusoidal light curves, and that
first or second overtone pulsators should have asymmetric light curves.
This also is in direct contradiction with the assumption that,
like RR Lyrae stars, the asymmetric
light curves are indicative of fundamental mode pulsation, with sinusoidal
light curves present in first --and maybe also in second-- overtone
pulsators (e.g. Nemec et al. 1995, McNamara 1995).


\section{The Evolutionary Status of $\delta$~Scuti Stars in the Bulge Fields}

The reddening in the bulge fields
is very inhomogeneous.  From the VR photometry it is convenient to use the
reddening independent magnitude $W_V$, defined as $W_V =  V - 3.97 \times 
(V - R)$, which assumes a standard extinction law for the bulge fields. 
Most of the $\delta$~Scuti stars
in our sample belong to the Galactic bulge;
their magnitudes peak at $W_V = 15.9$, which places them at about 8 kpc.

The $\delta$~Scuti are core hydrogen burning stars located inside the instability
strip, on the main sequence or just above the main sequence
(Breger 1995, Rodriguez et al. 1994). For a population older than about $10^{10}$ yr,
the instability strip lies bluewards of, and it is brighter than
the main sequence turn-off.  Thus, in old populations,
the pulsating variables in the instability strip must occur in blue stragglers.
It is now well established that SX~Phe stars in globular clusters
are blue stragglers (Nemec et al. 1994, 1995).

\vskip 4.08truecm
\begin{minipage}{4cm}
\includegraphics{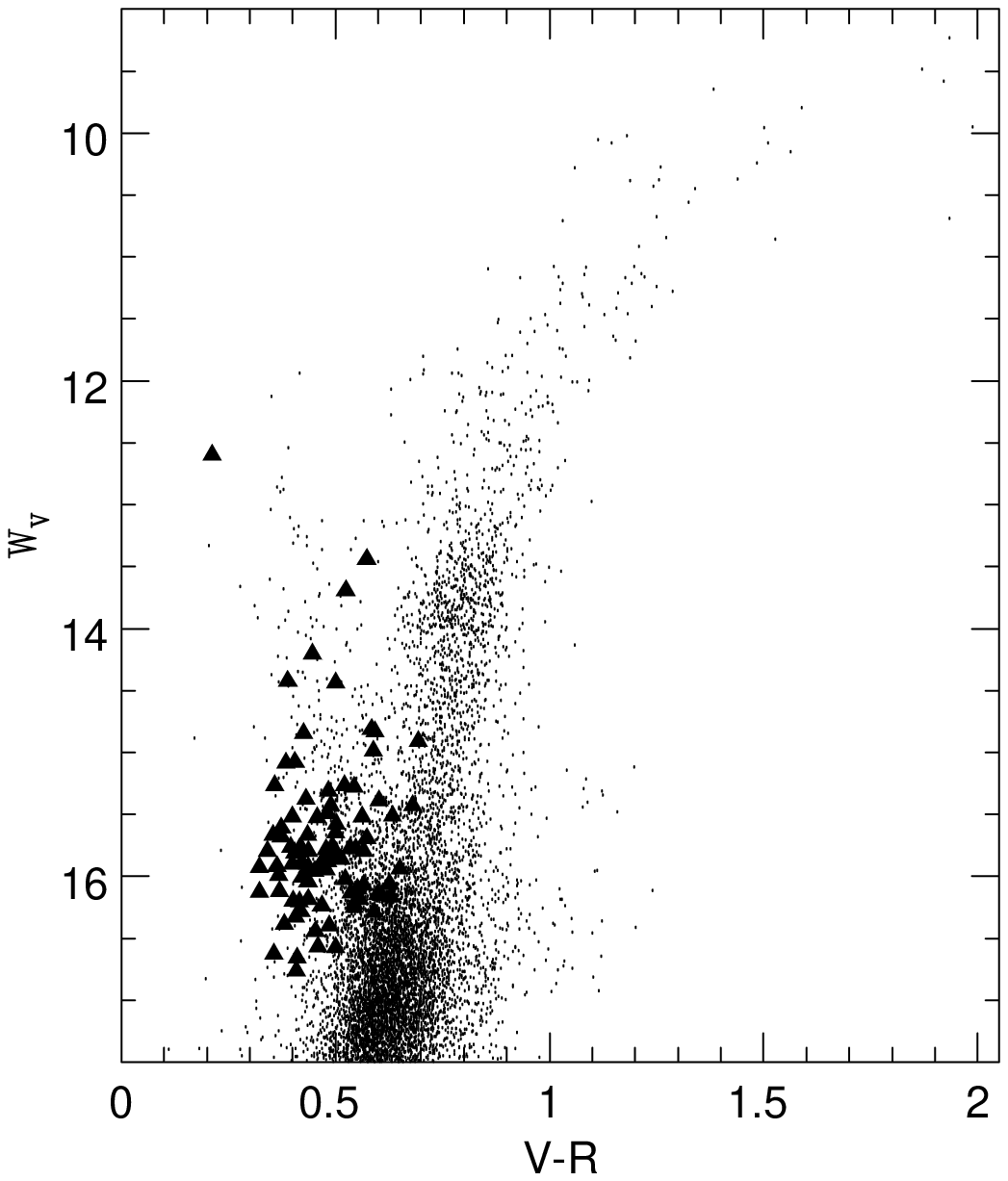}
\includegraphics{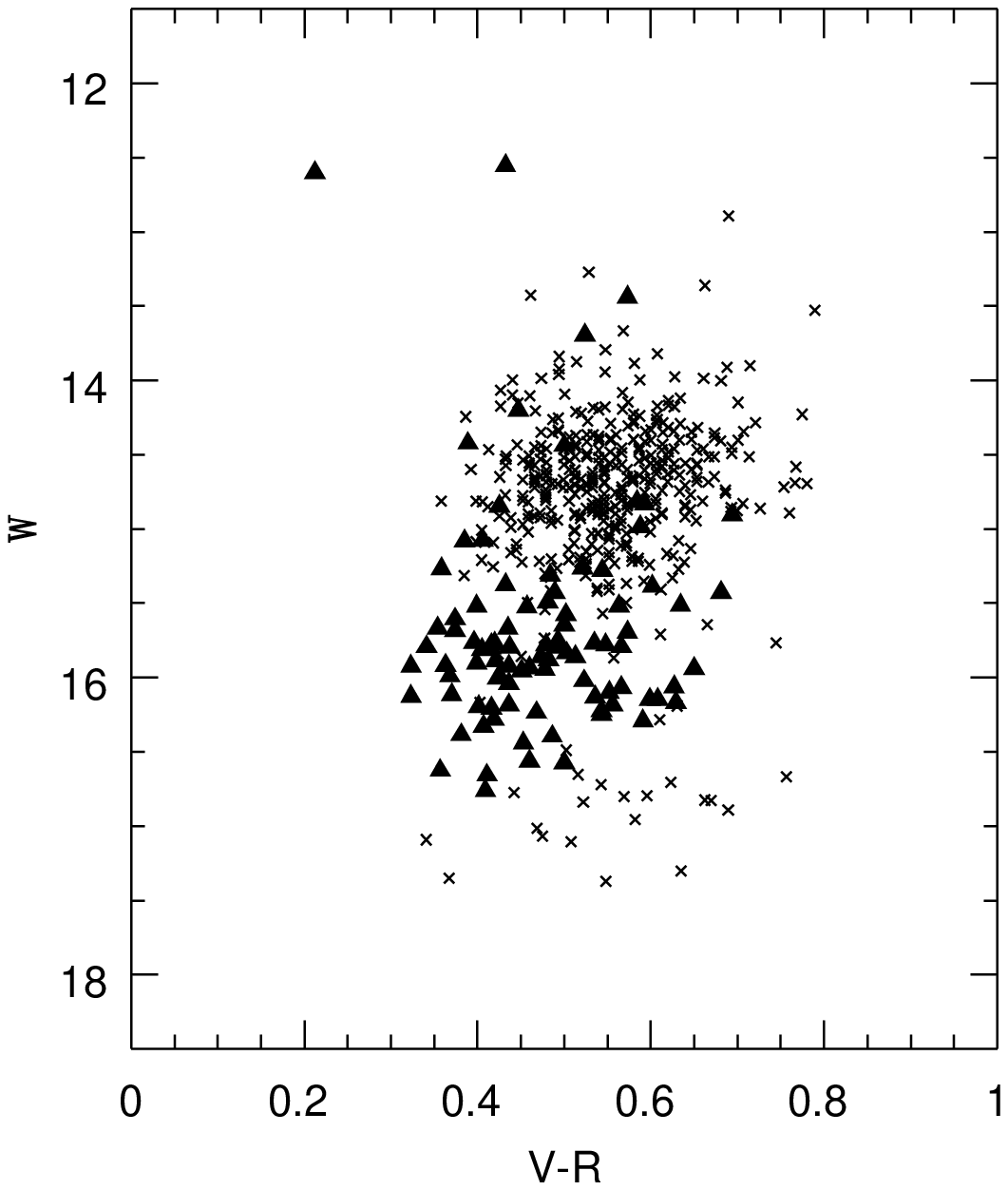}
\end{minipage}
\smallskip{}
\begin{center}
{\small Fig. 3 - Color-magnitude diagrams of bulge $\delta$ Scuti stars, 
compared with Baade's
window stars (left panel), and bulge RR Lyrae stars (right panel).}
\end{center}


Are the bulge $\delta$ Scuti stars also blue stragglers?
In order to answer this question,
Figure 3a shows the MACHO $W_V$ $vs$ $V-R$
color-magnitude diagram of Baade's window (only 5000 stars
are plotted). This color-magnitude diagram shows a prominent red giant
branch (RGB), a red horizontal branch (or RGB clump) at
$W_V = 13.5 - 14$, and the blue disk main sequence with $V-R < 0.5$.
The bulge turn-off would be located at the faint limit
of our photometry in this field. Note, however, that we reach the
bulge turn-off in other MACHO fields which are less crowded than Baade's window.
The $\delta$~Scuti stars with $0.08^d < P < 0.20^d$ from all the bulge fields are
also plotted (filled triangles) in Figure 3a.  From this Figure we conclude that
their colors and reddening-independent magnitudes
are consistent with bulge blue stragglers. This is confirmed also by the
location of $\delta$~Scuti stars (triangles) with respect to bulge RR Lyrae type ab
(crosses, from Alcock et al. 1997) in the color magnitude diagram shown in Figure 3b.

\section{The Distribution along the Line-of-sight}

Figure 4a shows the $W_V$ magnitude distribution of $\delta$ Scuti and RR Lyr
stars. The $\delta$ Scuti distribution is very 
peaked, with $FWHM = 0.6 \pm 0.05 $ mag.
This argues for $\delta$~Scuti stars pulsating in a single
dominant mode.  The observed magnitude scatter is consistent with the
line-of-sight depth of the Galactic bulge, and it would be larger if
pulsators in two or more different modes are represented.
Also note that the observed FWHM is slightly smaller
than that of RR Lyrae type ab stars. This means that the metallicity dependence
of the magnitude $\Delta M_V / \Delta [Fe/H]$ 
is smaller, either because the RR Lyrae have a wider
metallicity distribution, or because their spatial distribution is slightly
different. 

\vskip 3.08truecm
\begin{minipage}{3cm}
\includegraphics{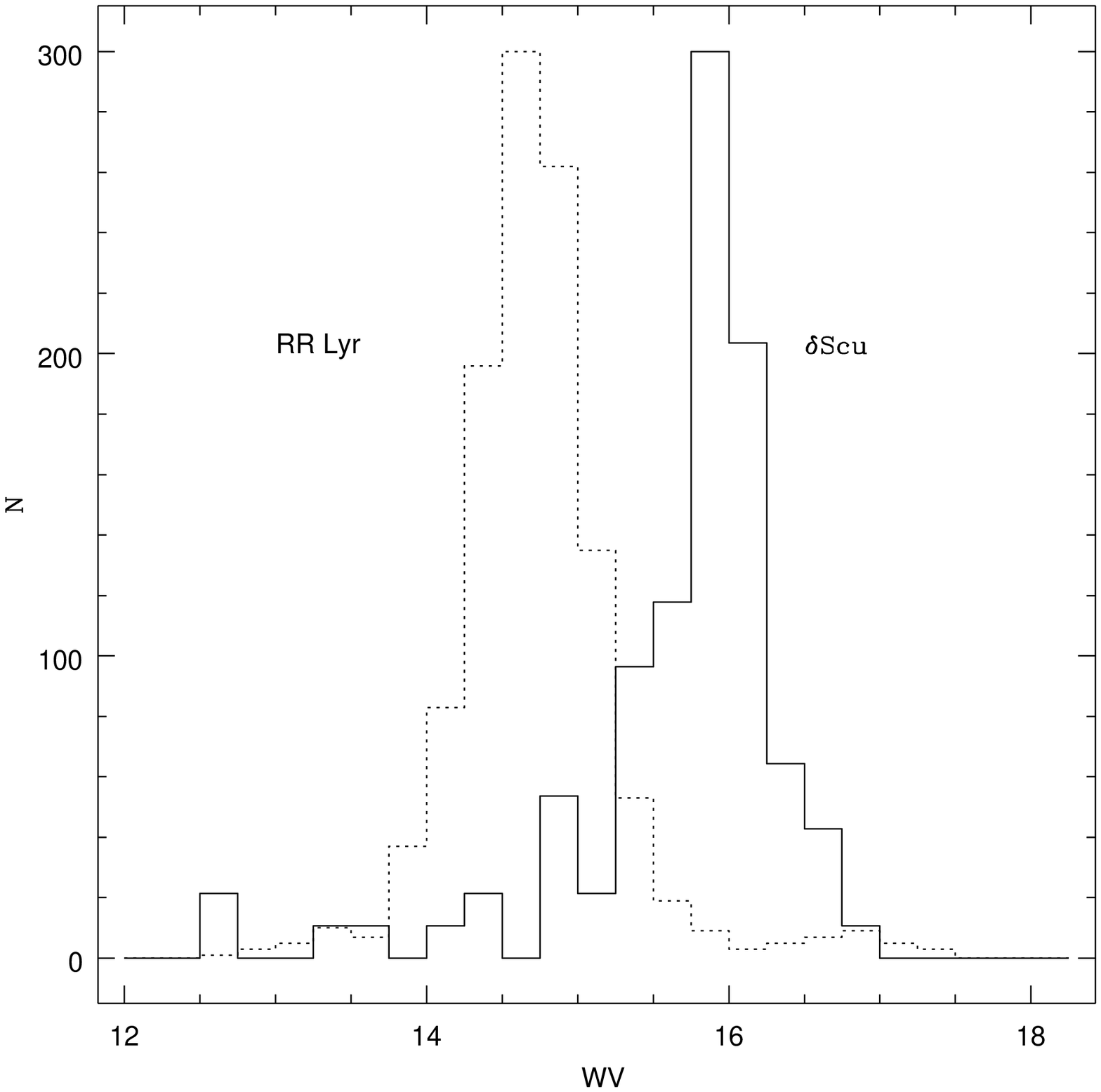}
\includegraphics{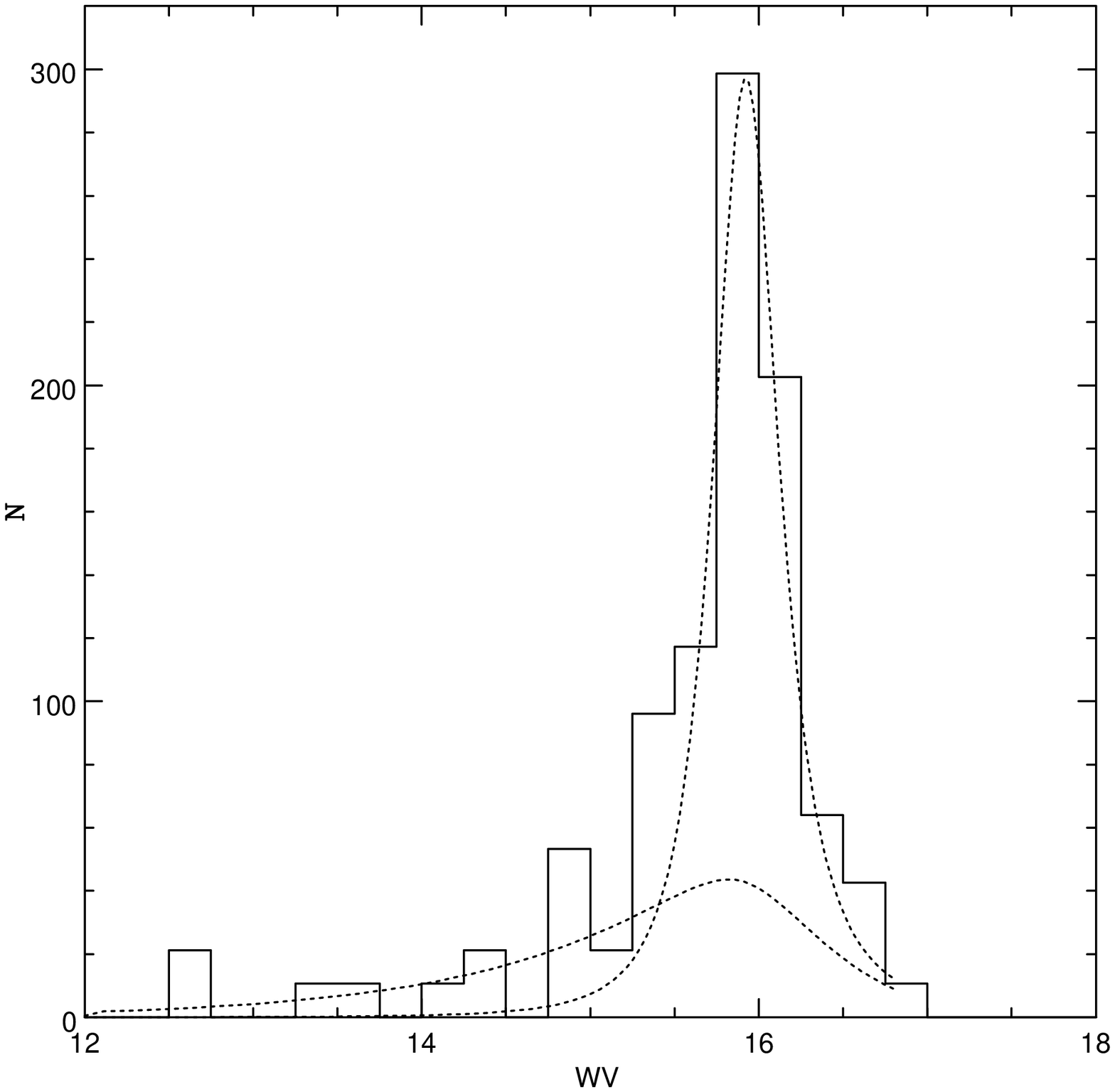}
\end{minipage}
\smallskip{}
\begin{center}
{\small Fig. 4 - Luminosity distribution of $\delta$~Scuti stars (arbitrary normalization).}
\end{center}


We assume that the line of sight
distributions of $\delta$~Scuti and RR~Lyrae stars are similar.
In the sky, these stars are concentrated towards the inner fields, as expected if
they follow the bulge radial density profile rather than the disk, which
would show a more uniform, shallower distribution.

In order to decide if the stars in the present sample belong to
the disk or the bulge, one can assume that all $\delta$~Scuti stars have
roughly the same magnitude, and compute the expected magnitude distribution
for these
two Galactic components integrated along the line of sight.
Figure 4b shows the observed magnitude
distribution, along with the line of sight density distribution expected for
the bulge and the disk (upper and lower dotted curves, respectively). 
The disk is assumed to be a double exponential of
the form $\rho_d \propto exp(-r/h_r) exp(-z/h_z)$,
with scale-height $h_z = 0.3$ kpc, and scale-length $h_r = 3$ kpc,
and the bulge is assumed to be approximated by a power law of the form
$\rho_b \propto r^{-n}$,
with $n = 3.5$. The counts have been normalized arbitrarily.
If all the $\delta$~Scuti stars from Table 1 were disk stars,
the FWHM of their observed magnitude distribution would be $\sim 2$ mag,
as seen in Figure 4b.  We conclude that the majority of the stars in our
sample are bulge $\delta$~Scuti stars.
However, about 10\% of the $\delta$~Scuti stars  shown in Figure 4
are significantly brighter than the mean. These could be $\delta$~Scuti stars
in the foreground disk, 
or merely blends.

\section{Conclusions}

We have analyzed a sample of
$\delta$~Scuti variable stars within a narrow period range,
$0.08^d < P < 0.20^d$, identified in the MACHO
bulge database.
The colors and magnitudes of these stars are consistent with them being
bulge blue stragglers.
A comparison of the observed light curves with the recent theoretical models
of Bono et al. (1997) suggests
that the present sample consists mainly of second overtone pulsators. 

We note that adopting the P--L--Z relations of Fernie (1992), McNamara (1995), or
Nemec et al. (1995), different mean absolute magnitudes can be obtained.
Distances measured using these $\delta$~Scuti stars are presently very uncertain,
because of the unknown metallicity dependence
and pulsational stages. 
The determination of an improved P--L relation for $\delta$~Scuti
stars, both from the theoretical and observational sides, would be very useful. 
For example, based on the tightness of the magnitude distribution of the bulge 
$\delta$~Scuti stars, when a firm P--L relation is established,
they could yield an independent distance to the Galactic center
as accurate as that measured using RR Lyrae stars. 


We would then like to stress the need for more a more complete exploration
of the parameter space (models with different 
masses, temperatures and luminosities).
%
Spectroscopy  of the present sample is also needed in order to determine
the abundances and masses of these stars, and to decide if
$\delta$~Scuti stars are the mere metal-rich extension of SX~Phe stars.

\acknowledgements{We would like to thank T. Bedding for useful suggestions,
and to all the organizers for a great meeting.  We are always very 
grateful for the skilled support by the technical staff at MSO.
Work at LLNL is supported by DOE contract W7405-ENG-48.
Work at the CfPA is supported NSF AST-8809616 and AST-9120005.
Work at MSSSO is supported by the Australian Department of Industry,
Technology and Regional Development. }

\end{document}